\documentclass[pra,aps,showpacs,showkeys,groupedaddress,superscriptaddress,twocolumn]{revtex4-1}

\usepackage[utf8x]{inputenc}
\usepackage{color}
\usepackage{bbm} 
\usepackage{amsfonts,amsmath,amssymb,stmaryrd}
\usepackage{graphicx}
\usepackage{subfigure}  % use for side-by-side figures
\usepackage{hyperref}
\usepackage{epsfig}
\usepackage{mathrsfs}
\usepackage{verbatim}
\usepackage{centernot}

\usepackage{array}

\usepackage{bm}	% bold in math mode

\begin{document}
\title{Finite-size corrections to quantized particle transport in topological charge pumps}

\author{Rui Li}
\affiliation{Department of Physics and Research Center OPTIMAS, University of Kaiserslautern, 67663 Kaiserslautern, Germany}
\author{ Michael Fleischhauer}
\affiliation{Department of Physics and Research Center OPTIMAS, University of Kaiserslautern, 67663 Kaiserslautern, Germany}

\begin{abstract}
We investigate the quantization of adiabatic charge transport in the insulating ground state of finite systems. Topological charge pumps are used in experiments as an indicator of topological order. In the thermodynamic limit, the transport can be related to a topological Berry phase and is thus strictly quantized. This is no longer true for finite systems. We derive finite-size corrections to the transport for both non-interacting and interacting systems and relate them to analytic properties of the single- and many-body Berry curvature.
We find that they depend on the details of experimental realizations of the pumps. While they can be non-negligible even in large systems, a
proper choice of the pumping protocol can suppress these corrections.
\end{abstract}

\date{\today}
\maketitle

%%%%%%%%%%%%%%%%%%%%%%%%%%%%%%
\section{introduction}
%%%%%%%%%%%%%%%%%%%%%%%%%%%%%%

Charge transport in electronic devices is usually associated with dissipation and heat production. 
Topological pumping, first introduced by Thouless \cite{Thouless-PRB-1983}, provides a robust and
controllable alternative for mesoscopic electronics  with minimal dissipation \cite{Brouwer-PRB-1998,Switkes-Science-1999,Makhlin-PRL-2001}. 
In such a topological or Thouless pump, an adiabatic cyclic variation of parameters leads to a strictly quantized transport in an insulating state of non-interacting fermions, which is related to an integer topological invariant, the Chern number. The concept can be generalized to interacting systems and the quantized transport survives moderate disorder \cite{Niu-JPhysA-1984,Avron-PRL-1985}. It is also closely related to the theory of polarization developed in the early 1990s \cite{King-Smith-1993,Ortiz-1994,Resta-1994,Resta-PRL-1998}. 
Topological pumping does not rely  on interaction effects such as Coulomb blockade
and can be observed for neutral particles as recently demonstrated with ultra-cold atoms \cite{Lohse-NatPhys-2015,Nakajima-NatPhys-2016}.
Imposing further symmetries, such as time-reversal symmetry, it is also possible to construct topological pumps for spins without net transport of charge
\cite{Shindou-JPhysSocJpn-2005,Fu-Kane-PRB-2006,Meidan-PRB-2010,Schweizer-arxiv-2016}, which has interesting applications in spintronics.

Charge or spin transport in an adiabatic Thouless pump is quantized, however,
only in the thermodynamic limit of infinite system size \cite{Thouless-PRB-1983}. The demand for size reduction in information technology (IT) makes it necessary to understand and minimize size-related deviations from quantized transport in topological pumps. This is the aim of the present paper.
Corrections to quantization of topological transport had been discussed before, e.g. in \cite{Rossini-PRB-2013}, and were attributed to
the finiteness of the critical gap when the system size is finite. Since there is a critical gap-closing in finite systems with periodic boundary conditions, this argument does not generally apply. 
The origin of these deviations is rather the discreteness of the momentum eigenmodes associated with finite systems. We show that the corrections decrease, first, polynomially, and then exponentially with system size L. The corresponding characteristic length scale $\xi$ can be related to analytic properties of the single- or many-body energy spectrum. In the case of non-interacting fermions, $\xi$ is determined by the width of the instantaneous single-particle band structure, and the transport properties can be optimized by a proper choice of the path of the Thouless pump in parameter space keeping this width as small as possible at all times.

As a specific example, we first discuss the simplest non-interacting topological charge pump, the Rice-Mele model at half filling. It describes fermions hopping on a one-dimensional lattice with staggered on-site energies and alternating hopping amplitudes \cite{Rice-PRL-1982,Wang-PRL-111,Xu-PRA-96}. We determine the 
characteristic length scale $\xi$ analytically and verify it with numerical results. We then consider one-dimensional models with interactions. 
Specifically, we discuss the super-lattice Bose Hubbard model  (SLBHM) \cite{SLBHM,Grusdt-PRL-2013} at half filling which is a bosonic analog of the Rice-Mele model and possesses a non-degenerate many-body ground state, but generalizations to other models including those with artificial dimensions are possible \cite{Zeng-PRL-115,Taddia-PRL-118}. Finally, we discuss the extended
superlattice Bose Hubbard model (E-SLBHM) at quarter filling, which has a two-fold degenerate ground state and a fractional topological charge \cite{ExtSLBHM,Grusdt-PhD}. As a consequence, a single
cycle of the adiabatic pump leads to a transport of only half a particle. Using time-evolving block
decimation (TEBD) simulations \cite{TEBD}, we show the exponential scaling of the corrections to the quantization of the particle transport. Although the present discussion is focusing on charge pumps, it can be straightforwardly generalized to spin pumps.

%%%%%%%%%%%%%%%%%%%%%%%%
\section{non-interacting fermions}
\label{sect1}
%%%%%%%%%%%%%%%%%%%%%%%%

We first discuss one-dimensional topological insulators of non-interacting fermions on a lattice with
period $a=1$ and finite number $L$ of unit cells, which is described by a single-particle Hamiltonian $H$. 
For simplicity, we restrict ourselves to one-dimensional band insulators, but the generalization to higher spatial dimensions is straightforward. 
Due to discrete translational invariance, the crystal momentum is conserved and can be restricted to the first Brillouin zone $q\in\{-\pi,\pi\}$ ($\hbar =1$). 
In a finite system with periodic boundary conditions, the crystal momentum takes on discrete values $q_j = 2\pi j/L -\pi$, for $j=1,2,...,L$. It is convenient to introduce the momentum-shifted Hamiltonian $H(q) = e^{-i q\hat x} \, H\,  e^{i q\hat x}.$
The eigenfunctions of $H(q)$ are cell-periodic Bloch functions $u_{nq}(x) = {\rm e}^{-iq x}\psi_{nq}(x) $, where the index $n$ denotes the $n$th Bloch band. 

We now assume that the parameters of the Hamiltonian are varied in time with period $T$, i.e., 
$H(t) = H(t+T)$, and that the system remains in a gapped state at all times. If the parameter variation is sufficiently slow and
encircles a gap-closing point, there can be an adiabatic
charge (spin) transport. As shown by Thouless {\it{et al.}} \cite{TKKN-PRL-1982,Thouless-PRB-1983}, this transport is strictly quantized
in the thermodynamic limit $L\to\infty$, and can be related to an integer topological invariant.  We will now revisit this
derivation.

The instantaneous (adiabatic) eigenstates of $H(q,t)$ are ${\rm e}^{-iq\hat{x}} \vert u_n(q,t)\rangle$. In order to determine the adiabatic current and the transported charge, we need to consider 
corrections to theses states up to the first order in the rate of change of the Hamiltonian. Assuming a non-degenerate ground state
$\vert u_0(q,t)\rangle$ with a finite energy gap, we find in the lowest order of time-dependent perturbation theory
\begin{equation}
\vert \psi_0(q)\rangle = \vert u_0(q)\rangle + i\sum_{n\ne 0} \frac{\vert u_n(q)\rangle\langle u_n(q)\vert {\partial_t u}_0(q)\rangle}{\varepsilon_n(q)-\varepsilon_0(q)}.
\end{equation}
Here, $\varepsilon_n(q)$ are the instantaneous eigenenergies and we dropped the overall dynamical phase factor which will be canceled later on as well as the dependence on $t$ for notational convenience. The single-particle velocity operator
$\hat v = -i[\hat x,H]$ reads in the momentum-shifted frame $\hat v(q)=e^{-iq\hat x} \hat v e^{iq \hat x} = \partial H(q)/\partial q$, which yields in the state
$|\psi_0(q)\rangle$:
\begin{eqnarray}
&&v_0(q) =\langle\psi_0(q)\vert\, \hat  v\, \vert \psi_0(q)\rangle \\
&&\enspace= \frac{\partial \varepsilon_0(q)}{\partial q} +i \sum_{n\ne 0}\left(
\frac{\langle u_0\vert\partial_q H(q) \vert u_n\rangle\langle u_n\vert \partial_t u_0\rangle}{
\varepsilon_n(q)-\varepsilon_0(q)} - c.c.\right)\nonumber\\
&&\enspace = \frac{\partial \varepsilon_0(q)}{\partial q} +i \left(\Bigl\langle\frac{\partial u_0}{\partial t}\Bigl\vert \frac{\partial u_0}{\partial q}\Bigr\rangle-
\Bigl\langle\frac{\partial u_0}{\partial q}\Bigl\vert \frac{\partial u_0}{\partial t}\Bigr\rangle
\right).\nonumber 
\end{eqnarray}
In the last step we have used that $\langle u_0(q) \vert \partial_q H(q) \vert u_n(q)\rangle = \langle \partial_q u_0(q)\vert u_n(q)\rangle  \big(\varepsilon_0(q) -\varepsilon_n(q)\big)$, which follows directly from the eigenvalue equation of the momentum-shifted
Hamiltonian.

In an insulating state, we have to add the contributions of all occupied momentum modes to obtain the total current. In particular, for systems with only the lowest Bloch band occupied, $J_L = \frac{1}{L}\sum_{j=1}^L v_0(q_j)$. The total charge (particle number) $Q_L$, transported
in a period $T$, is then given by the integral of the current. Taking into account 
that the time-independent Hamiltonian does not support a current when summing over all quasi-momenta of a
band, one finds
\begin{eqnarray}
&& Q_L = \int_0^T\!\!\!  dt \,  J_L \equiv  \int_0^T\!\!\!  dt \, \frac{1}{L} \sum_{j=1}^{L}  \Omega_0(q_j,t)\label{eq:charge}  \\
&& =  \int_0^T \!\!\! dt \frac{1}{L}\sum_{j=1}^{L} i\left(
\Bigl\langle\frac{\partial u_0}{\partial t}\Big\vert \frac{\partial u_0}{\partial q}\Big\rangle-\Big\langle\frac{\partial u_0}{\partial q}\Big\vert \frac{\partial u_0}{\partial t}\Big\rangle
\right)\!\!\bigg|_{q=q_j}.
\nonumber
\end{eqnarray}
where $\vert u_0\rangle \equiv \vert u_0(q)\rangle$ and $\Omega_0(q_j,t)$ is the Berry curvature of the $n=0$ Bloch band.
In the thermodynamic limit, $L\to\infty$,
the sum in eq.(\ref{eq:charge}) can be replaced by an integral
$\frac{1}{L}\sum_{j=1}^{L} f_j  = \int_{-\pi}^{\pi} \frac{dq}{2\pi} f(q)$, and one obtains an integral
over a closed surface of a torus
\begin{equation}
Q_L= -i \int_0^T\!\!\! dt\int_{-\pi}^{\pi} \!\frac{dq}{2\pi}
 \left(
\Bigl\langle\frac{\partial u_0}{\partial q}\Bigl\vert \frac{\partial u_0}{\partial t}\Bigr\rangle -
\Bigl\langle\frac{\partial u_0}{\partial t}\Bigl\vert \frac{\partial u_0}{\partial q}\Bigr\rangle\right),
\end{equation}
which is an integer number \cite{Thouless-PRB-1983}. 

For a finite system, however, the sum over lattice momenta can not be replaced by an integral. As
a consequence, the transported charge is no longer quantized.
In the following, we will discuss the deviation of the transported charge $Q$ from its thermodynamic limit $Q_L$: $\Delta Q_L = Q_L -Q$.
In most relevant cases, the Berry curvature $\Omega_0(q)$ is analytic in the whole Brillouin zone, i.e. there exists a strip $(-\pi,\pi) \times (-c,c)$
in the extension of the Brillouin zone to the complex $q$ plane where $\Omega_0(q,t)$ is analytic and its
derivatives exist to all orders and they are periodic in $q$. 
While generically the difference between an integral and its approximation by a finite sum decreases only polynomially in $1/L$, it has been shown
in \cite{Javed-PRSc-2013}  that it can scale exponentially 
for integrals of periodic functions and 
is determined by the value of $c=c(t)$:
\begin{equation}
\vert \Delta Q_L\vert \le  \int_0^T\!\! dt\,  \frac{2 M e^{-c(t) L}}{1-e^{-c(t) L}}.
\label{eq:charge-correction}
\end{equation}
Here $M$ is a bound on $\vert\Omega_0(q,t)\vert$ within the first Brillouin zone.
For small systems, $\vert\Delta Q_L\vert$ scales polynomially as $1/L$ and turns over to
an exponential scaling for large $L$.
The characteristic length $\xi$ beyond which the charge transport is approximately quantized
 is determined by the values of $1/c(t)$ along the parameter path of the pump. If the parameter path is chosen
 such that $c(t)=c = \xi^{-1}$ is constant in time, a simple exponential scaling emerges. 
 For systems sizes $L$ smaller than $\xi$ the transport is no longer integer quantized.
 
We will now show that $\xi$ 
is determined by the curvature of the band structure $\varepsilon_n(q)$. To see this we write the Berry curvature in the form
\begin{equation}
\Omega_0(q) = i \sum_{n\ne 0}\left(
\frac{\langle u_0\vert\partial_q H(q) \vert u_n\rangle\langle u_n\vert \partial_t u_0\rangle}{
\varepsilon_n(q)-\varepsilon_0(q)} - c.c.\right).\label{eq:I}
\end{equation}
%
%where we have suppressed the $q$-dependence in the Bloch wave functions for notational simplicity.
Eq.(\ref{eq:I}) shows that $\Omega_0(q)$ attains a pole in the complex $q$-plane when the energy gap closes for a complex value 
$q= q^\prime + i q^{\prime\prime}$. Most importantly,
in the flat-band limit where $\varepsilon(q) =$ const, the Berry curvature is analytic in the whole complex
plane, and thus the adiabatic charge transport is strictly quantized irrespective of system size. Thus choosing
a topological pump which operates as close as possible to the flat-band limit will support strictly quantized charge
transport even for very small systems.

It is interesting to note at this point that while the transported charge is strictly quantized only in the thermodynamic limit,
a related quantity, the winding of the (electric) polarization, is quantized for arbitrary system size.
Within the theory of polarization, King-Smith and Vanderbilt \cite{King-Smith-1993}
and Resta \cite{Resta-PRL-1998} showed that in the
thermodynamic limit $L\to \infty$,
the adiabatic particle (charge) current $J_L$ coincides  with the time derivative of the many-body polarization 
\begin{equation}\label{eq:Polarization}
J_{L\to\infty} =\frac{\partial}{\partial t} P,\qquad
P = \frac{1}{2\pi} \textrm{Im}\ln \left\langle e^{i \frac{2\pi}{L}\hat X}\right\rangle,
\end{equation}
where $\hat X = \sum_{j=1}^N \hat x_j$ is the total position operator of all $N$ particles.
The polarization winding after a full period of an adiabatic charge pump, $\Delta P =\int_0^T dt\, \partial_t P$, 
is given by the Chern number of the pump and is thus integer quantized (for lattice constant $a=1$). (Note that $2\pi P$, eq.(\ref{eq:Polarization}), is the phase of a complex exponential. As such, its winding after going through a full cyclic variation in parameter space is trivially integer-quantized modulo $2\pi$.)
% It is interesting to note  that while the transported charge is only strictly quantized in the thermodynamic limit $L\to\infty$, the polarization winding $\Delta P$ is always integer valued.

\begin{figure}[htb]
	\begin{center}
	\includegraphics[width=\columnwidth]{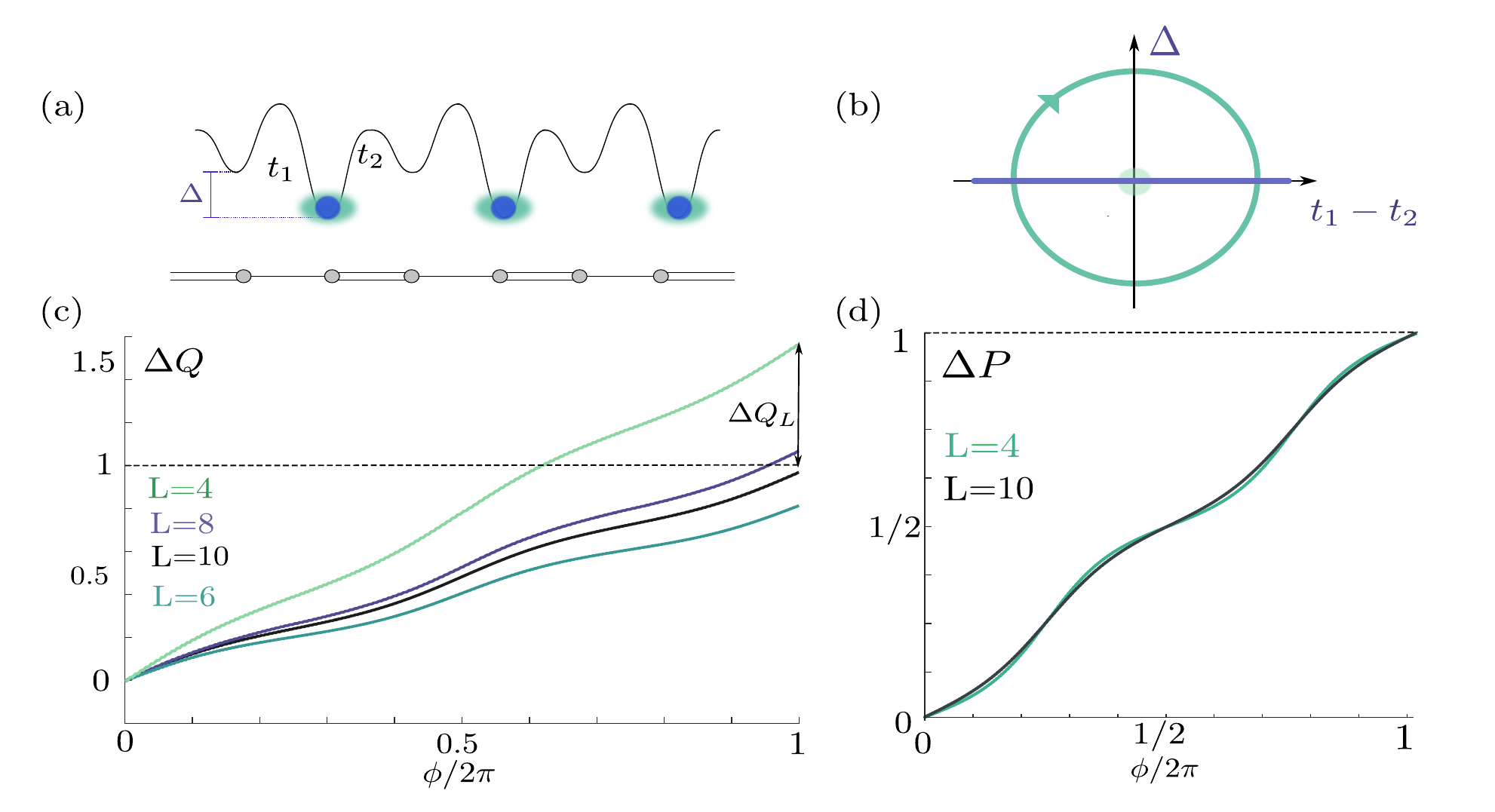}
	\end{center}
	\caption{(a) Rice-Mele model. (b) Sketch of the protocol for charge pumping in the parameter space of the Rice-Mele model. (c) Transported charge as a function of angle $\phi$ in the $\Delta - (t_1-t_2)$ plane for different values of $L$. The path for the pump is parameterized as $t_1=1-0.5 \cos\phi$, $t_2 =1+0.5 \cos\phi$, and $\Delta= 2/\sqrt{3} \sin\phi$.  (d) The same for the polarization. One recognizes the strictly integer-valued winding of $P$ for all values of $L$, while the particle transport is only quantized in the thermodynamic limit $L\to\infty$.}
	\label{fig:RM-pump}
\end{figure}

In the following, we will illustrate our findings for a simple topological model of non-interacting fermions, the Rice-Mele model
\cite{Rice-PRL-1982}, shown in Fig.\ref{fig:RM-pump} (a).
Here, fermions move along a one-dimensional lattice with alternating hopping amplitudes $t_1,t_2 \ge 0$ and a staggered on-site energy offset $\Delta$. In second quantization, the Hamiltonian reads
\begin{eqnarray}
H &=& -t_1\sum_{j,\textrm{even}} c_j^\dagger c_{j+1}
 -t_2\sum_{j,\textrm{odd}} c_j^\dagger c_{j+1} +h.a.\nonumber\\
 &&
-\Delta \sum_j (-1)^j c_j^\dagger c_j,\label{eq:RM-Hamiltonian}
\end{eqnarray}
where $c_j, c_j^\dagger$ are fermionic annihilation and creation operators at lattice site $j$.
Since the unit cell consists of two sites, the single-particle energy spectrum has two bands
$\varepsilon_{\pm}(q) = \pm \varepsilon(q)$
\begin{equation}
\varepsilon(q) =\sqrt{\Delta^2 +(t_1+t_2 e^{iq})(t_1+t_2 e^{-iq})}.
\end{equation}
The band gap closes for $\Delta=0$ and $t_1=t_2$. For $\Delta=0$, the Rice-Mele model reduces to the
Su-Schrieffer-Heeger model \cite{SSH-PRL-1979}. At half filling, the latter possesses two different topological phases protected by inversion symmetry, which differ in their Zak (or Berry) phase by $\pi$. The two phases cannot be
smoothly connected without closing the energy gap or breaking the inversion symmetry. However, introducing the staggered potential allows one to
adiabatically connect the two phases. Performing a closed loop in the parameter space of $\Delta$ and $t_1-t_2$ encircling the origin leads to a quantized transport of a single charge (in the thermodynamic limit). The charge transport can be related to an effective Chern number.
 
Extending lattice momenta to the complex $q$-plane, i.e., $q=q^\prime +iq^{\prime\prime}$, one finds that there is a closing of the energy gap
for $q^\prime = \pi$ and $\cosh\left(q^{\prime\prime}\right) =A=\frac{\Delta^2 +t_1^2 +t_2^2}{2 t_1 t_2}$.
This yields for the characteristic length 
\begin{equation}
\xi^{-1} \simeq \ln\left(A+\sqrt{A^2-1}\right)= \ln\left(\frac{1+\Delta \varepsilon}{1-\Delta\varepsilon}\right).
\label{eq:xi}
\end{equation}
Here $\Delta\varepsilon=\varepsilon(q)_\textrm{min}/\varepsilon(q)_\textrm{max}$ is the energy gap relative to the total energy width. 
One recognizes that in the limit of flat bands, where $t_1t_2=0$ at all times and consequently $\varepsilon(q) =\mathrm{const}_q$, the characteristic length vanishes, $\xi =0$.
In this limit the adiabatic charge transport is strictly quantized for any system size $L$.
Away from this limit there are exponential corrections to the transported charge, see Fig.\ref{fig:RM-pump} (c),
while the polarization winding $\Delta P$ always
attains an integer value, see Fig.\ref{fig:RM-pump} (d). For small relative energy gaps the characteristic length can become rather large ($\xi\sim 1/\Delta\varepsilon$) leading to substantial
corrections even for rather large systems.

%%%%%%%%%%%%%%%%%%%%%%%%%%%%%%%%%%%%%%%%%
%%%%%%%%%%%%%%%%%%%%%%%%%%%%%%%%%%%%%%%%%%%

\begin{figure}[htb]
	\begin{center}
	\includegraphics[width=\columnwidth]{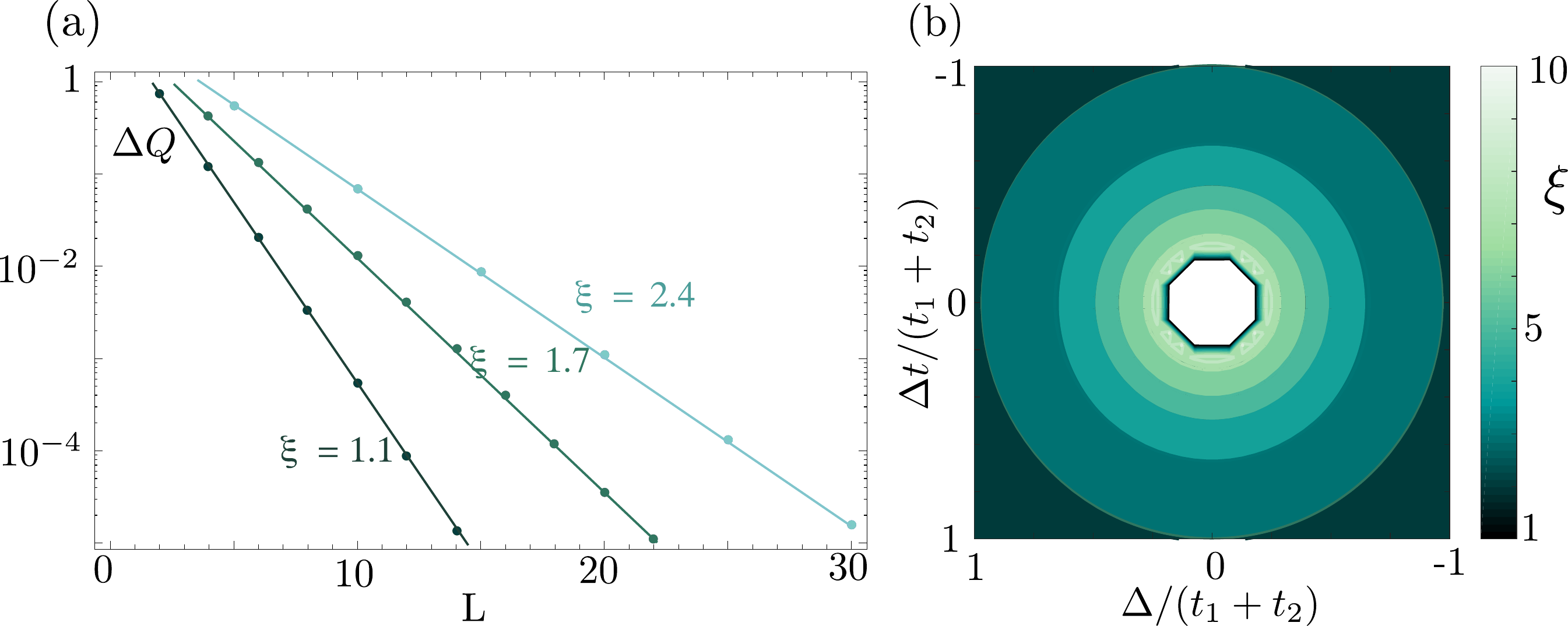}
	\end{center}
	\caption{(a) Transported charge as a function of $L$.  Shown are numerical results obtained from eq.(\ref{eq:charge}) (dots) and exponential
	fits (full lines) for parameter paths $t_{1,2} = 1 \mp  \sqrt{ (A-1)/(A+1)} \cos\phi $, and $\Delta = \sqrt{2 (A-1)} \sin\phi $, and $A= 1.1, 1.2, 1.5$ (from the top).
	The fitted length scales fit to the bounds obtained from eq.(\ref{eq:xi}).
	(b) Contour plot of characteristic length scale of 
	finite-size corrections for the Rice-Mele model as a function of $\Delta t=t_1-t_2$ and $\Delta$, eq.(\ref{eq:xi}), cut at $\xi = 10$. (The octagonal shape of the inner part is due to finite numerical precision.)
	}
	\label{fig:size-scale}
\end{figure}

We have verified the system-size dependence according to eq.(\ref{eq:charge-correction})
numerically, see Fig.\ref{fig:size-scale}(a).
There we have plotted
the transported charge as a function of system size obtained 
%either from a full time-dependent simulation of the Thouless pump for small system sizes $(L=2,\dots,7)$ or
from evaluating the finite sum, eq.(\ref{eq:charge}), for a larger range of system sizes $(L=2,\dots, 30)$. The parameter path of the pump has been chosen in such a way that
$c(t) = \mathrm{const}_t$.  One clearly recognizes the predicted exponential scaling and the extracted characteristic length fit to the estimates given in eq.(\ref{eq:xi}). Fig.\ref{fig:size-scale}(b)
shows the characteristic length $\xi$ as function of $\Delta t= t_1-t_2$ and $\Delta$. 
The closer the parameter path encircles the critical point, the larger value $\xi$ takes.
In this regime, 
finite-size corrections to the particle transport can become non-negligible even for rather large systems.

%%%%%%%%%%%%%%%%%%%%%%%%%%%%%%%%%%%%%%%%%
%%%%%%%%%%%%%%%%%%%%%%%%%%%%%%%%%%%%%%%%%%%

%%%%%%%%%%%%%%%%%%%%%%%%%%%
\section{interacting systems}
%%%%%%%%%%%%%%%%%%%%%%%%%%%

%%%%%%%%%%%%%%%%%%%%%%%%%%%
\subsection{Thouless pump for non-degenerate ground states}
%%%%%%%%%%%%%%%%%%%%%%%%%%

The above discussion can be extended to interacting many-body systems or systems with disorder. 
Let us first consider the case of an interacting lattice model with a non-degenerate ground state.
The transport in a system of size $L$ 
upon time-periodic changes of the Hamiltonian can be calculated in a similar way as in section \ref{sect1}, replacing the single-particle wave functions by the many-body eigenstates $\vert \Phi_n\rangle$.
Assuming a finite energy gap between the many-body ground state $\vert \Phi_0\rangle $ and the excited states, the transported charge in a period $T$ can be expressed as
\begin{equation}
Q_L = -\frac{i}{L} \int_0^T\!\!\! dt\, \sum_{n\ne 0} \frac{\langle \Phi_0\vert P\vert \Phi_n\rangle\langle \Phi_n\vert \partial_t \Phi_0\rangle}{E_0-E_n}+ c.c.
\label{eq:QL}
\end{equation}
Here $P=\sum_{i=1}^N p_i= -i\sum_{i=1}^N\partial/\partial{x_i}$ denotes the total momentum of all $N$ particles. 

Niu and Thouless have shown  that in the thermodynamic
limit $N,L\to \infty$ with $N/L=$ const., the transported charge in an insulating state can be related to an integral of an appropriate Berry curvature over a closed surface and 
thus is integer quantized \cite{Niu-JPhysA-1984}. To see this they considered the ground state of the $N$-particle Hamiltonian
with twisted boundary conditions: i.e. 
\begin{equation}
\Phi(x_1,\dots,x_j+L,\dots,x_N) = e^{i\beta} \Phi(x_1,\dots,x_j,\dots,x_N)
\end{equation}
for all  $j\in\{1,...,N\}$. Here $\beta = \alpha L$ is a continuous parameter that can be varied from $-\pi$ and $\pi$. 
A canonical transformation $\vert \Psi\rangle = \exp\{-i\alpha \sum_j x_j\} \vert \Phi\rangle$ transforms the problem 
to one with periodic boundary conditions and new Hamiltonian 
$H(\alpha) = e^{-i\alpha \hat X} H e^{i\alpha \hat X}$.
%, where $\hat X = \sum_{j=1}^N \hat x_j$ is the total position operator of all $N$ particles.
$H(\alpha)$ contains a gauge potential $\alpha$, i.e. all particle momenta  $p_j=-i\partial/\partial x_j$ are replaced by $\tilde p_j=-i\partial/\partial x_j +\alpha$, i.e.
$P\to\tilde P$. 
The corresponding many-body eigenstates and eigenenergies 
become $\alpha$ dependent, i.e. $\vert \Psi_n(\alpha)\rangle$ and $E_n(\alpha)$. 

Let us now consider the pumped charge in the ground state of $H(\alpha)$.
Making use of $\langle \Phi_n\vert \tilde P \vert \Phi_0\rangle =\langle \Phi_n\vert \partial_\alpha\vert \Phi_0\rangle (E_0-E_n)$
one finds
\begin{eqnarray}
Q(\alpha) &=& -\frac{i}{L} \int_0^T\!\!\! dt\, \left( \sum_{n\ne 0} \langle \partial_\alpha\Phi_0\vert \Phi_n\rangle\langle \Phi_n\vert \partial_t \Phi_0\rangle - c.c.\right)\nonumber\\
&=& -\frac{i}{L} \int_0^T\!\!\! dt\, \left(\Bigl \langle \frac{\partial\Phi_0}{\partial\alpha}\Bigr\vert 
\frac{\partial\Phi_0}{\partial t}\Bigr\rangle - \Bigl \langle \frac{\partial\Phi_0}{\partial t}\Bigr\vert 
\frac{\partial\Phi_0}{\partial \alpha}\Bigr\rangle\right).
\end{eqnarray}
Averaging over all values of $\alpha$ in $\{-\pi/L,\pi/L\}$ yields
\begin{eqnarray}
\bar Q &=&\int_{-\pi}^\pi \frac{d\beta}{2\pi} Q_L = \frac{L}{2\pi} \int_{-\pi/L}^{\pi/L}\!\!\! d\alpha \, Q_L(\alpha) \\
&=&-\frac{i}{2\pi} \int_0^T\!\!\! dt \int_{-\pi/L}^{\pi/L}\!\!\! d\alpha \, \left(\Bigl \langle \frac{\partial\Phi_0}{\partial\alpha}\Bigr\vert 
\frac{\partial\Phi_0}{\partial t}\Bigr\rangle - \Bigl \langle \frac{\partial\Phi_0}{\partial t}\Bigr\vert 
\frac{\partial\Phi_0}{\partial \alpha}\Bigr\rangle\right).\nonumber
\end{eqnarray}
$\bar Q$ is an integral of the many-body Berry curvature $\Omega_0(\alpha,t)$ over a closed surface and is thus integer quantized.
Niu and Thouless argued that for $L\to\infty$
\begin{equation}
Q_L \equiv Q(\alpha=0)= \bar Q
\end{equation}
and the transported charge becomes
integer quantized.

To obtain the finite-size corrections we note that the difference between $\bar Q$ and $Q_L=Q(0)$ is just the error of the
mid-point approximation of the integral, which is given by a similar expression as in eq.(\ref{eq:charge-correction}) \cite{Javed-PRSc-2013}
\begin{equation}
\left\vert Q_L -\bar Q\right\vert \le \int_0^T \!\! \! dt \, 
\frac{2 M e^{-1/\zeta}}{1-e^{-1/\zeta}} = \int_0^T \!\!\! dt \, \frac{2 M e^{-L/\xi}}{1-e^{-L/\xi}} 
\label{eq:many-body-correction}
\end{equation}
Here $\Omega_0(\beta)$ is analytic in $\beta\in \{-\pi,\pi\} \times \{-\zeta^{-1},\zeta^{-1}\}$ or equivalently in $\alpha\in 
 \{-\pi/L,\pi/L\} \times \{-\xi^{-1},\xi^{-1}\}$. Thus the finite-size corrections to adiabatic charge transport are determined by the
 analytic properties of the many-body Berry curvature corresponding to the Hamiltonian $H(\alpha)$ in a complex-valued gauge field $\alpha=\alpha^\prime + i \alpha^{\prime\prime}$. The characteristic length $\xi$ of finite-size corrections can be obtained from the closure of the many-body gap $\Delta E(\alpha)$ for a complex
 $\alpha$. 
  
\begin{figure}[htb]
	\begin{center}
		\includegraphics[width=\columnwidth]{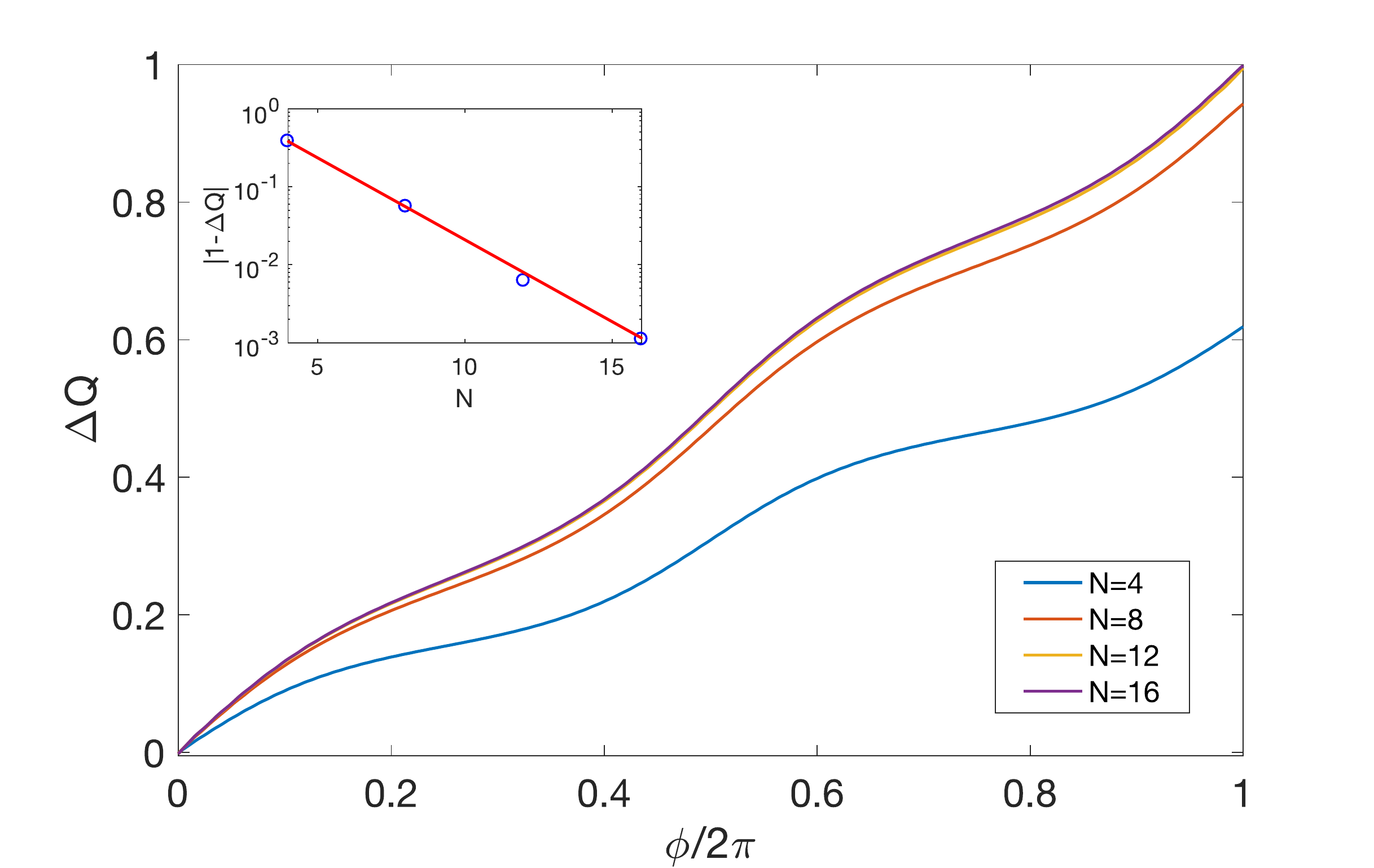}
	\end{center}
	\caption{Transported charge in the superlattice Bose Hubbard model at half filling as function of $\phi$ for different $N$.	
	The path for the pump is parameterized as $t_{1,2}=1\mp 0.5345 \cos\phi$, $\Delta=1.2649\sin\phi$, and onsite interaction $U=153.4508$.
	The inset shows the deviation of the transported charge after one cycle from unity.}
	\label{fig:SLBHM}
\end{figure}

 Estimating $\xi$ requires analytic knowledge of the many-body  gap which is in general rather involved. We will thus restrict ourselves in the following to
 verifying the exponential size-scaling 
 % and determining the dependence of the characteristic length scale on system parameters
  numerically. To this end we use
 TEBD simulations \cite{TEBD} with periodic boundary conditions. Specifically we consider the bosonic analog of the Rice-Mele model, the 
 super-lattice Bose Hubbard model \cite{SLBHM}.  The Hamiltonian is identical to (\ref{eq:RM-Hamiltonian}), with bosonic rather than fermionic operators and with
 an additional term $H_1 =\sum_j (U/2) n_j(n_j-1)$ describing onsite repulsion with strength $U>0$. In the hard-core limit, realized for $U\gg  t_1, t_2, \vert \Delta\vert$
 the model can be mapped to the Rice Mele model. In Fig.\ref{fig:SLBHM} we show the dependence of the transported charge on the number of sites $N$. The results verify the exponential scaling.
 
 %%%%%%%%%%%%%%%%%%%%%%%%%%%%%%%%%%%%%%%%%

 %%%%%%%%%%%%%%%%%%%%%
 \subsection{Thouless pump for degenerate ground states and $U(n)$ Berry phase}
 %%%%%%%%%%%%%%%%%%%%%
 
Due to interactions, the ground state can spontaneously break the discrete translational symmetry of the underlying model and
multiple degenerate ground states can exist. In such a case a topological pump can transfer one of the ground states into the other states and multiple
pump cycles are required to return to the original bulk state. The topological invariant describing such a quantized charge pump in the thermodynamic limit
is then a $U(n)$ Berry phase, where $n$ is the degree of degeneracy. The above discussion can straightforwardly be generalized to this case. The only difference is that the Berry curvature is integrated over the time of a cycle returning the broken-symmetry bulk state to itself, which is a multiple of the 
time period of the underlying many-body Hamiltonian.

\begin{figure}[htb]
	\begin{center}
	\includegraphics[width=\columnwidth]{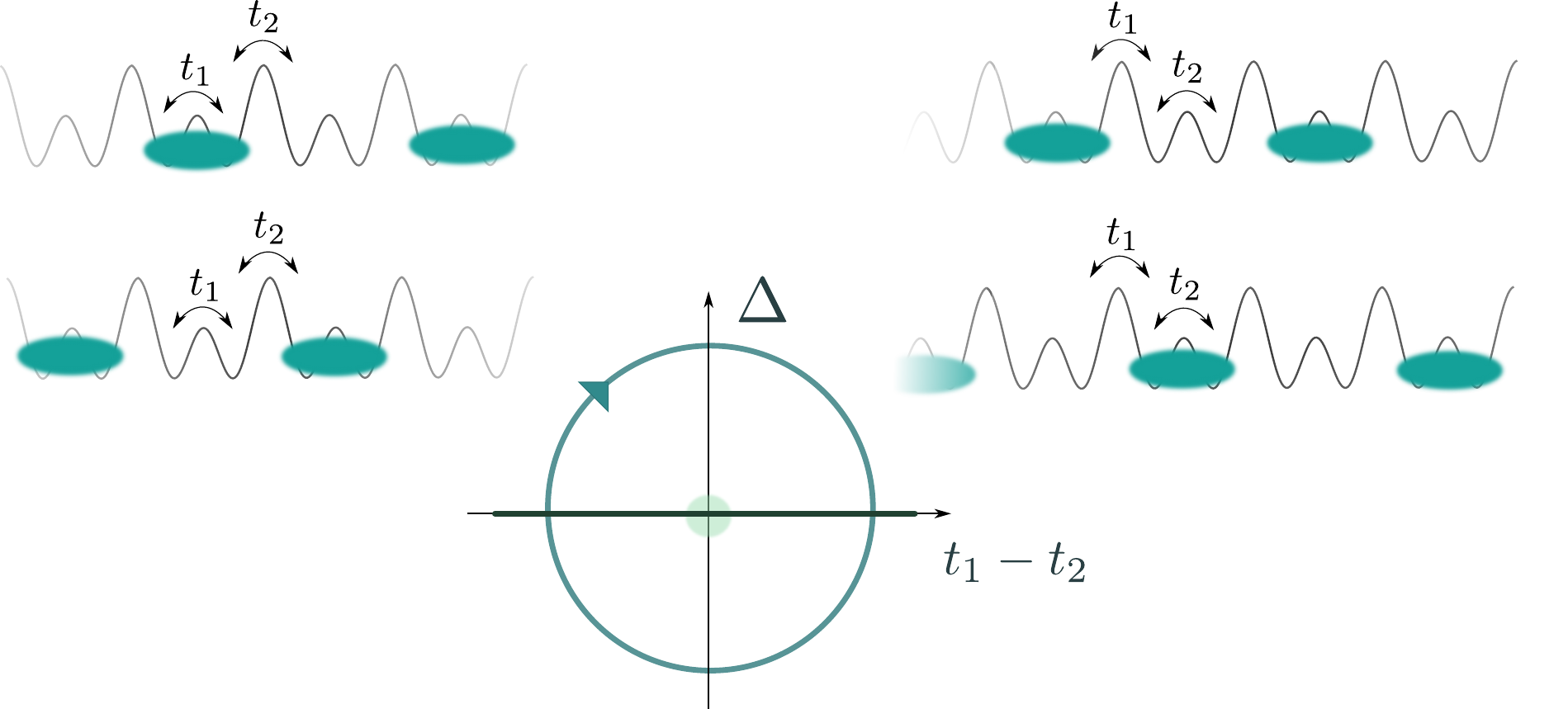}
		\end{center}
	\caption{Schematics of the Thouless pump for $\rho=1/4$. The ground state is two-fold degenerate and two pump cycles are needed to return to the initial bulk state.	}
	\label{fig:ExtSLBHM}
\end{figure}

\begin{figure}[htb]
	%\begin{center}
	%\includegraphics[width=0.495\columnwidth]{fig5_a.pdf}
	%\includegraphics[width=0.49\columnwidth]{fig5_b.pdf}
	\centering
	\subfigure[]{\includegraphics[width=0.493\columnwidth]{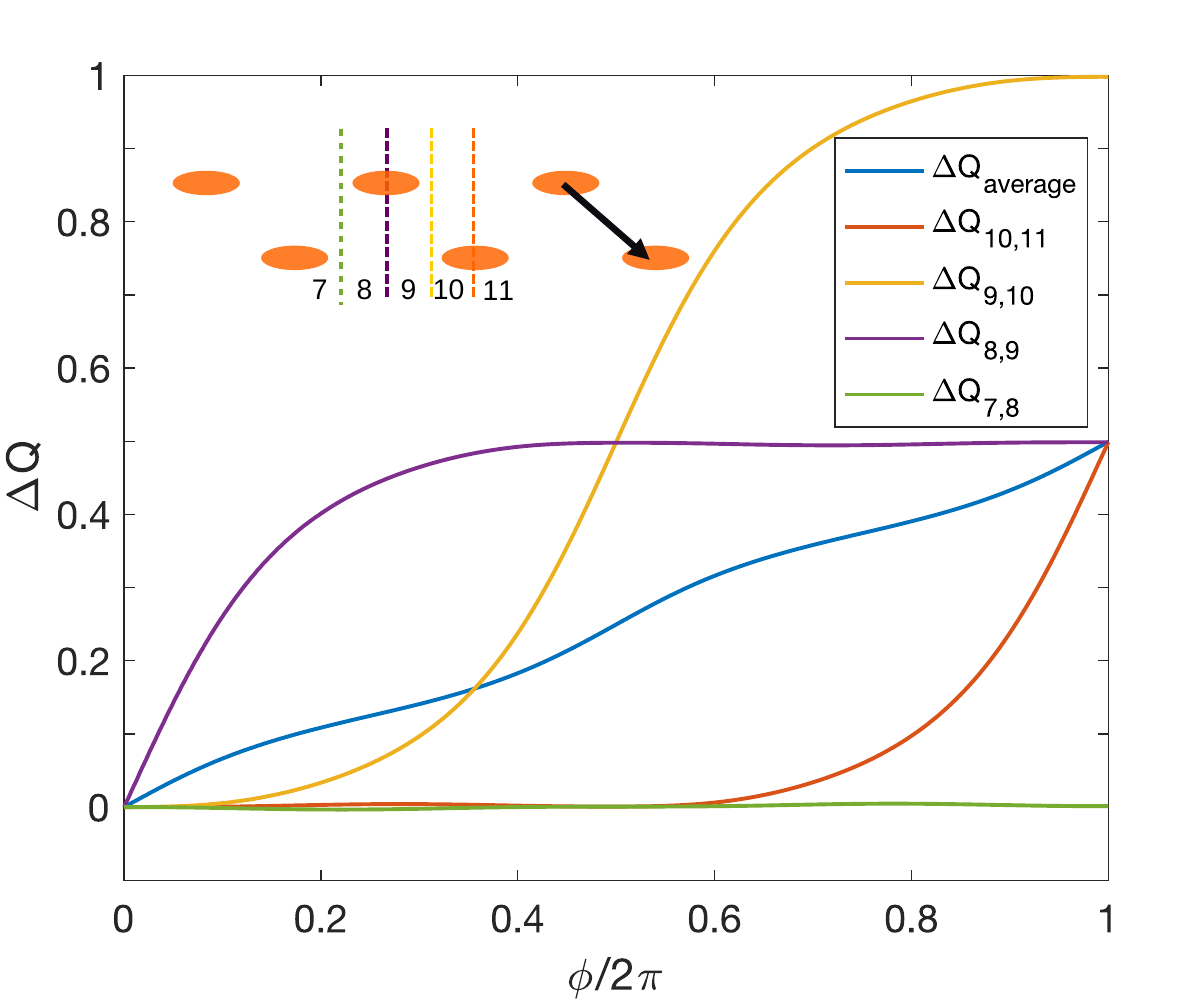}}
    %\qquad
    \subfigure[]{\includegraphics[width=0.495\columnwidth]{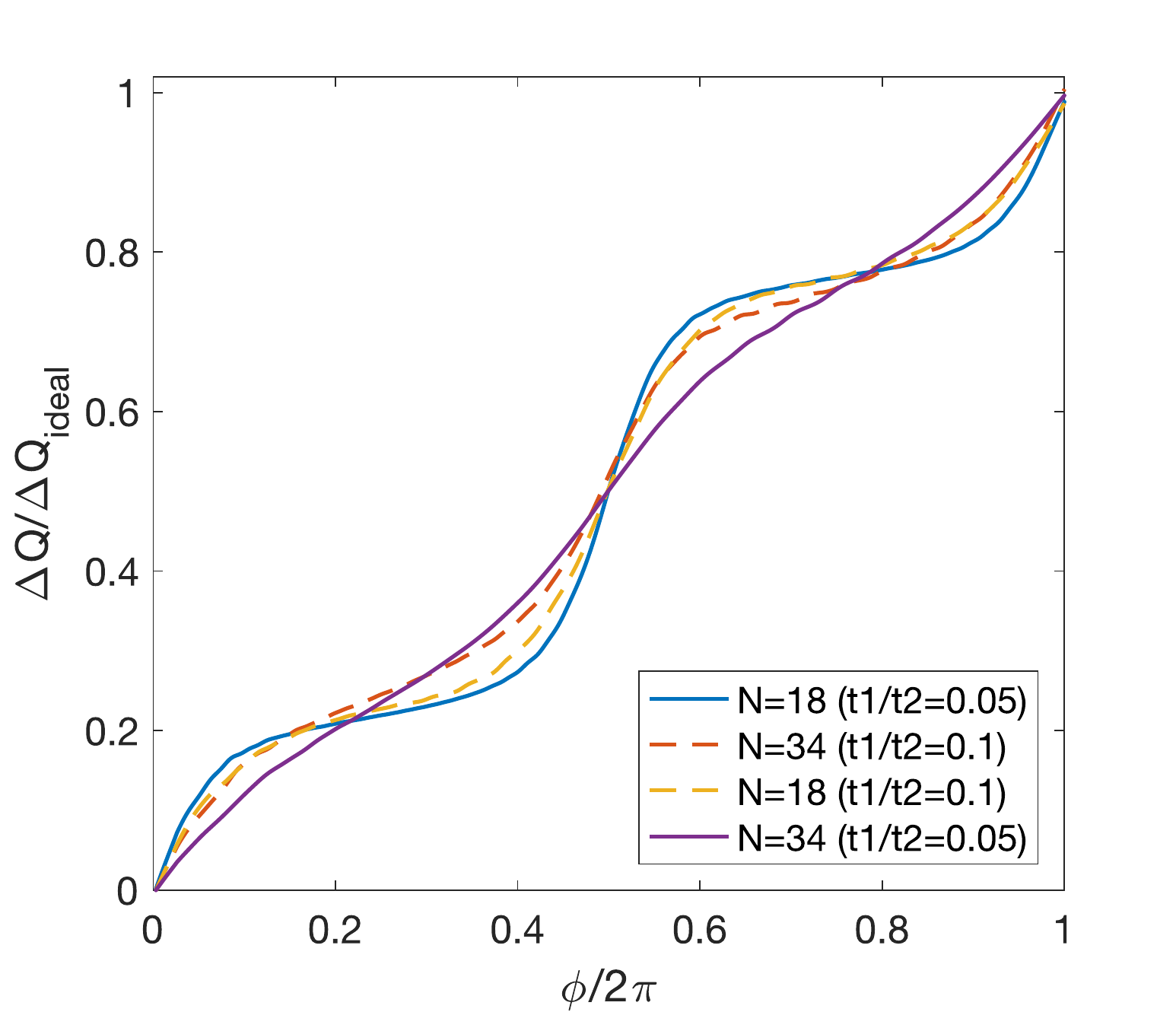}}
	\caption{(a) Transported charge across successive bonds in the unit cell of the E-SLBHM (see inset) as function of $\phi$ in the atomic limit. The pump cycle 
	is parameterized $t_{1,2}= \frac{1}{2}\bigl(1\mp \cos(\phi)\bigr)$, $\Delta=\sin(\phi)$. Here $N=18$ lattice sites and open boundary conditions are considered.
	Note that due to the two-fold degeneracy two pump cycles are needed for an integer particle transport.
	(b) Averaged transported charge $\Delta Q$ normalized to $\Delta Q_\textrm{ideal}$ in a system with open boundary conditions. Dashed lines show results for $N=18$ and 			$N=34$ for $t_{1,2}= \frac{1}{2}\bigl(1.1\mp 0.9\cos(\phi)\bigr)$, $\Delta=1.56 \sin(\phi)$. The deviations of transported charges from their ideal values are $|\Delta Q-\Delta Q_\textrm{ideal}|=0.0067$ and $0.001$ respectively. 
	Full lines correspond to $N=18$ and $N=34$ for $t_{1,2}= \frac{1}{2}\bigl(1.05\mp 0.95\cos(\phi)\bigr)$, $\Delta=0.95 \sin(\phi)$. Here, $|\Delta Q-\Delta Q_\textrm{ideal}|=0.0038$ and $0.0012$ respectively.  $\Delta Q_\textrm{ideal}$ takes into account the contributions of the edge states at the beginning and the end of the pump cycle, $\Delta Q_\textrm{ideal}=0.5294$ for  $N=18$ and $0.5151$ for $N=34$.	
	Parameters for both figures are $V_1=4$, $V_2=2$ and $U=40$.
	} 
	\label{fig:ExtSLBHM-charge}
\end{figure}

We will now illustrate this for the example of the  extended SLBHM \cite{ExtSLBHM,Zeng-PRB-2016}. This model 
is similar to the SLBHM but contains in addition nearest and next-nearest neighbor interactions $V_1$ and $V_2$ respectively.
 \begin{eqnarray}
 && H = -t_1\sum_{\textrm{even}} a_j^\dagger a_{j+1}
 -t_2\sum_{\textrm{odd}} a_j^\dagger a_{j+1} -\Delta \sum_j (-1)^j n_j\nonumber\\
 &&\quad + \sum_j\left(\frac{U}{2}n_j(n_j-1) + V_1 n_j n_{j+1} + V_2 n_j n_{j+2}\right)
 \end{eqnarray}
Here $n_j=a_j^\dagger a_j$. For sufficiently large values of $U$ and $V_{1,2}$ this model has Mott-insulating (MI) ground states with fractional
filling, which spontaneously break the translational symmetry of the superlattice.  MI phases exist for fractional fillings $\rho=1/4, 1/3, 1/2 ...$. In the following we will consider the $\rho=1/4$ MI state which is doubly degenerate. For $\Delta =0$ the Hamiltonian is inversion symmetric and possesses four distinct ground-state phases illustrated in Fig.\ref{fig:ExtSLBHM} for the atomic limit 
($U\gg V_1>V_2 \gg$ min$[t_1,t_2]$). These phases can be distinguished by their behavior under inversion at a fixed bond and by the Zak phase with respect to that
bond, which defines a topological quantum number \cite{Grusdt-PhD}. A Thouless pump transfers the bulk state into itself only after two cycles and is associated with
a $U(2)$ Berry phase. As a consequence the pumped charge in a single cycle averaged over all bonds and in the thermodynamic limit is $1/2$. This is illustrated in Fig.\ref{fig:ExtSLBHM-charge} (a), where we show numerical results obtained by TEBD. 
Since TEBD simulations are very difficult for periodic boundary conditions we here choose open boundary conditions. To avoid the influence of the edges we
calculate the transported charge only at bonds in the center of the chain. We consider conditions where the distance to the boundaries is much larger 
than the localization length of the edge states and where this length is smaller than the anticipated characteristic length scale
of the transport. We also make sure that the Thouless pump does not lead to transitions into higher bands at the edges.
Due to the density-wave character of the ground state the transported charge differs for different bonds but averages to about $0.5$. In Fig.\ref{fig:ExtSLBHM-charge} (b) we illustrate the charge transport for different system sizes normalized to the ideal values that take into account that in a single pump cycle the occupied edge state moves from the left to the right.
As expected $\Delta Q/\Delta Q_\textrm{ideal}$ approaches unity with increasing system size and when the difference of the tunneling rates is larger.

%%%%%%%%%%%%%%%%%%%%%%%%%%%%%%%%%%%%%%%%%
%%%%%%%%%%%%%%%%%%%%%%%%%%%%%%%%%%%%%%%%%%%

 %%%%%%%%%%%%%%%%%%%%%%%%%%%%%%%%%%%%%%%%%
%%%%%%%%%%%%%%%%%%%%%%%%%%%%%%%%%%%%%%%%%%%

%%%%%%%%%%%%%%%%%%%%%%%%
\section{summary and outlook}
%%%%%%%%%%%%%%%%%%%%%%%

Adiabatic topological transport of charge in insulating ground states is in general not quantized in finite systems. 
We derived an analytical upper bound to deviations from integer values for both non-interacting and interacting systems, which results from the discreteness of momentum space in finite systems and is determined by analytic properties of the Berry curvature. Specifically, we considered the Rice-Mele model as an example of a non-interacting model exhibiting topological order. Through dynamical simultations of a charge pump and direct evaluation of the finite sum eq.\eqref{eq:charge-correction}, we verified the exponential scaling of the corrections to the quantized particle transport, which agrees with analytic predictions. Furthermore, we investigated the same effect for the super-lattice Bose-Hubbard model and the extended super-lattice Bose-Hubbard model as examples of interacting systems. A slightly modified argument using the many-body wave-function can be made for the existence of exponential finite-size corrections in this case. However, the evaluation for the explicit expression of the characteristic length scale is only possible if 
analytic knowledge of the many-body gap exists. We verified the exponential scaling using TEBD. Our findings suggests that deviations can become non-negligible even for larger systems, which may explain small corrections on top of non-adiabatic contributions observed in recent experiments \cite{Lohse-NatPhys-2015}. On a more conceptual level, our findings highlight the difference between the winding of Berry (Zak) phase (or polarization) and quantized transport. While the former indicate the existence of topological order in systems of any size, adiabatic transport strictly shows topological order only for infinite systems. 

%
%\
%
%
%\centerline{\texttt{to be written}}
%
%\

%%%%%%%%%%%%%%%%%%%%%%%%
\subsection*{acknowledgement}
%%%%%%%%%%%%%%%%%%%%%%%%

Financial support by the Deutsche Forschungsgemeinschaft (DFG) through the SFB-TR 185 is gratefully acknowledged. The authors thank Dominik Linzner for invaluable support. M. F. would like to thank Chares-Edouard Bardyn and Sebastian Diehl
for stimulating discussions. This research 
was supported in part by the National Science Foundation under Grant No. NSF PHY-1125915.

%%%%%%%%%%%%%%%%%%%%%%%%%%%%%%%%%%%%%%%%%%%%%%%%%%%%%%%%%%%%%%%%%


\begin{thebibliography}{99}
\bibitem{Thouless-PRB-1983} D.J. Thouless, {\it Quantization of particle transport}, Phys. Rev. B \textbf{27}, 6083 (1983).
\bibitem{Brouwer-PRB-1998} P.W. Brouwer, {\it Scattering approach to parametric pumping}, Phys. Rev. B \textbf{58}, R10 135, (1998).
 \bibitem{Switkes-Science-1999} M. Switkes, C.M. Marcus, K. Campman, A.C. Gossard, {\it An adiabatic quantum electron pump}, Science \textbf{283}, 1905 (1999).
 \bibitem{Makhlin-PRL-2001} Yuriy Makhlin and Alexander D. Mirlin, \textit{Counting Statistics for Arbitrary Cycles in Quantum Pumps}
Phys. Rev. Lett. \textbf{87}, 276803 (2001).
\bibitem{Niu-JPhysA-1984} Q. Niu, and D.J. Thouless, {\it Quantised adiabatic charge transport in the preence of substrate disorder and many-body interactions}, J. Phys. A, \textbf{17}, 2453 (1984).
\bibitem{Avron-PRL-1985} Joseph E. Avron and Ruedi Seiler, \textit{Quantization of the Hall Conductance for General, Multiparticle Schrodinger Hamiltonians},
Phys. Rev. Lett. \textbf{54}, 259 (1985).
\bibitem{King-Smith-1993} R. D. King-Smith and David Vanderbilt, \textit{Theory of polarization of crystalline solids}, Phys. Rev. B \textbf{47}, 1651 (1993).
\bibitem{Ortiz-1994}  Gerardo Ortiz and Richard M. Martin, \textit{Macroscopic polarization as a geometric quantum phase: Many-body formulation}, Phys. Rev. B \textbf{49}, 14202 (1994).
\bibitem{Resta-1994} Raffaele Resta, \textit{Macroscopic polarization in crystalline dielectrics: the geometric phase approach}, Rev. Mod. Phys. 
\textbf{66}, 899 (1994)
\bibitem{Resta-PRL-1998} R. Resta {\it Quantum Mechanical Position Operator in Extended Systemy}, Phys. Rev. Lett. \textbf{80}, 1800 (1998)
\bibitem{Lohse-NatPhys-2015} M. Lohse, C. Schweizer, O. Zilberberg, M. Aidelsburger and I. Bloch, \textit{A Thouless quantum pump with ultracold bosonic atoms in an optical superlattice}, Nat. Phys. \textbf{12}, 350 (2015).
\bibitem{Nakajima-NatPhys-2016}  Shuta Nakajima,	Takafumi Tomita, Shintaro Taie,	Tomohiro Ichinose, Hideki Ozawa, Lei Wang, Matthias Troyer, and Yoshiro Takahashi, \textit{Topological Thouless pumping of ultracold fermions} Nat. Phys. \textbf{12}, 296  (2016).
\bibitem{Shindou-JPhysSocJpn-2005} Ryuichi Shindou, \textit{Quantum Spin Pump in S = $1/2$ Antiferromagnetic Chains - Holonomy of Phase Operators in sine-Gordon Theory -}, J. Phys. Soc. Jpn. \textbf{74}, 1214 (2005).
\bibitem{Fu-Kane-PRB-2006} Liang Fu and C.L. Kane, \textit{Time reversal polarization and a $Z_2$ adiabatic spin pump}, Phys. Rev. B \textbf{74}, 195312  (2006). 
\bibitem{Meidan-PRB-2010} Dganit Meidan, Tobias Micklitz, and Piet W. Brouwer, \textit{Optimal topological spin pump}, Phys. Rev. B \textbf{82}, 161303(R)  (2010). 
\bibitem{Schweizer-arxiv-2016} C. Schweizer, M. Lohse, R. Citro, I. Bloch, \textit{Spin pumping and measurement of spin currents in optical superlattices}, Phys. Rev. Lett. 117, 170405 (2016).
\bibitem{Rossini-PRB-2013} Davide Rossini, Marco Gibertini, Vittoria Giovannetti, and Rosario Fazio, {\it Topological pumping in the one-dimensional Bose-Hubbard model}, Phys. Rev. B \textbf{87}, 085131 (2013).
\bibitem{Rice-PRL-1982} M. J.Rice and E.J. Mele, \textit{Elementary Excitations of a linearly Conjugated Diatomic Polymer}, Phys. Rev. Lett. \textbf{49}, 1455 (1982).
\bibitem{Wang-PRL-111} L. Wang, M. Troyer, and X. Dai, \textit{Topological Charge Pumping in a One-Dimensional Optical Lattice}, Phys. Rev. Lett. \textbf{111}, 026802 (2013).
\bibitem{Xu-PRA-96} Z. Xu, Y. Zhang, and S. Chen, \textit{Topological phase transition and charge pumping in a one-dimensional periodically driven optical lattice}, Phys. Rev. A \textbf{96}, 013606 (2017).

\bibitem{SLBHM} P. Buonsante, V. Penna, and A. Vezzani, \textit{Fractional-filling loophole insulator domains for ultracold bosons in optical superlattices},
Phys. Rev. A \textbf{70}, 061603(R) (2004)
\bibitem{Grusdt-PRL-2013} F. Grusdt, M. Höning, and M. Fleischhauer,
\textit{Topological edge states in the one-dimensional super-lattice Bose-Hubbard model},
Phys. Rev. Lett. \textbf{110}, 260405 (2013).
\bibitem{Zeng-PRL-115} Tian-Sheng Zeng, Ce Wang, and Hui Zhai, \textit{Charge Pumping of Interacting Fermion Atoms in the Synthetic Dimension}, Phys. Rev. Lett. \textbf{115}, 095302 (2015).
\bibitem{Taddia-PRL-118} L.Taddia, E. Cornfeld, D. Rossini, L. Mazza, E. Sela, and R. Fazio \textit{Topological Fractional Pumping with Alkaline-Earth-Like Atoms in Synthetic Lattices}, Phys. Rev. Lett. \textbf{118}, 230402 (2017).
\bibitem{ExtSLBHM} F. J. Burnell, Meera M. Parish, N. R. Cooper, and S. L. Sondhi, \textit{Devil's staircases and supersolids in a one-dimensional dipolar bose gas}, Phys. Rev. B, \textbf{80},174519 (2009); Zhihao Xu, Linhu Li, and Shu Chen,
\textit{Fractional Topological States of Dipolar Fermions in One-Dimensional Optical Superlattices},
Phys. Rev. Lett. \textbf{110}, 215301 (2013).
\bibitem{Grusdt-PhD} F. Grusdt, PhD thesis, University of Kaiserslautern (2015).
\bibitem{TEBD} Guifre Vidal, \textit{Efficient Simulation of One-Dimensional Quantum Many-Body Systems},
Phys. Rev. Lett. \textbf{93}, 040502 (2004);
A. J. Daley, C. Kollath, U. Schollwock and G. Vidal, \textit{Time-dependent density-matrix renormalization- group using adaptive effective Hilbert spaces}, J. Stat. Mech., P04005 (2004).
\bibitem{TKKN-PRL-1982} D.J. Thouless, M. Komoto, M.P. Nightingale, and M. den Nijs, {\it Quantized Hall Conductance in a Two-Dimensional Periodic Potential}, Phys, Rev. Lett. \textbf{49}, 6083 (1982).
\bibitem{Javed-PRSc-2013} Mohsin Javed and Lloyed N. Trefethen, 
\textit{A trapezoidal rule error bound unifying the Euler-Maclaurin formula and geometric convergence for periodic functions},
Proc. Roy. Soc. A, \textbf{470}, 20130571 (2013), Corollary 3.1.
\bibitem{SSH-PRL-1979} W.P. Su, J.R. Schrieffer and A. J. Heeger, \textit{Solitons in Polyacetylene}, Phys. Rev . Lett. \textbf{42}, 1698 (1979).
\bibitem{Zeng-PRB-2016} Tian-Sheng Zeng, W. Zhu, and D. N. Sheng, \textit{Fractional charge pumping of interacting bosons in one-dimensional superlattice},
Phys. Rev. B \textbf{94}, 235139  (2016).

\end{thebibliography}
\end{document}